\documentclass[aps,prapplied,reprint,superscriptaddress,longbibliography,floatfix]{revtex4-2}
\usepackage{bm}
\usepackage{amsmath}
\usepackage{amsfonts}
\usepackage{amssymb}
\usepackage{graphicx}
\usepackage{CJK,color}  
\usepackage{xcolor}
\usepackage[braket, qm]{qcircuit}   
\usepackage[colorlinks,linkcolor=blue,anchorcolor=blue,citecolor=blue,urlcolor=blue]{hyperref}
\usepackage{dcolumn}

\begin{document}


\title{ Quadratic speed-up of quantum adiabatic evolution with engineered dissipation}
\begin{CJK}{UTF8}{gbsn}
\author{Yuanyang Zhou(周远洋)}
\affiliation{International Center for Quantum Materials, School of Physics, Peking University, Beijing 100871, China}

\author{Biao Wu(吴飙)}
\email{wubiao@pku.edu.cn}
\affiliation{International Center for Quantum Materials, School of Physics, Peking University, Beijing 100871, China}
\affiliation{Wilczek Quantum Center, Shanghai Institute for Advanced Studies, University of Science and Technology of China， Shanghai 201315, China}
\affiliation{Hefei National Laboratory, Hefei 230088, China}

\date{\today}

\begin{abstract}
Quantum adiabatic evolution is often slowed by nonadiabatic leakage near small spectral gaps. We propose an engineered dissipative protocol that continuously relaxes these excitations during the adiabatic evolution. In the ideal limit, a spectrally filtered jump operator induces downward transitions in the instantaneous energy eigenbasis while leaving the instantaneous ground state dark. An effective avoided-crossing analysis shows that in the regime where the engineered relaxation strength is much larger than the minimum gap, the runtime scaling can improve from the closed-system behavior $\mathcal{O}(\Delta^{-2})$ to $\mathcal{O}(\Delta^{-1})$. Finite-temperature upward transitions introduce a thermal error floor, the inverse-gap runtime estimate remains applicable when the thermal excited-state population at the minimum gap stays below a fixed fraction of the target error tolerance. Numerical results show improved ground-state preparation and robustness against weak upward-jump errors over closed-system annealing. We also discuss possible superconducting-circuit implementations using structured bosonic reservoirs and the microscopic consistency requirements for realizing the effective dynamics.
\end{abstract}

\maketitle
\end{CJK}
\section{Introduction}
Adiabatic quantum computation (AQC) and quantum annealing provide a natural framework for preparing low-energy states of many-body Hamiltonians and for solving optimization problems encoded in ground states~\cite{Farhi2000,Kadowaki1998,Das2008,AlbashLidar2018,Rajak2023}. In the ideal closed-system picture, the system is initialized in the ground state of a simple Hamiltonian and is then slowly transformed into the target Hamiltonian. If the evolution is sufficiently slow compared
with the inverse powers of the relevant spectral gaps, the system remains close to the instantaneous ground state. The central difficulty is that small avoided crossings generate nonadiabatic leakage from the instantaneous ground state to excited states. As a result, the runtime of closed-system adiabatic state preparation is often limited by the minimum gap, with the conventional estimate scaling as $T=\mathcal{O}(\Delta^{-2})$ for fixed target accuracy~\cite{Kato1950,AlbashLidar2018}.

Several approaches address this bottleneck. Local adiabatic schedules slow the evolution near small gaps~\cite{Roland2002QuantumSearch}, while parallel-eigenvalue paths maintain a constant gap in adiabatic quantum search~\cite{Daems2008Parallel}. Eigenstate filtering offers a digital route to ground-state preparation~\cite{ge2019faster,lin2020near} and has been combined with digitized adiabatic evolution using quantum singular value transformation (QSVT)-based filters~\cite{karacan2025filter}. Non-Abelian adiabatic mixing provides another coherent mechanism, exploring independent-set solutions within a degenerate ground-state manifold~\cite{WuYuWilczek2020}. Dissipation offers a complementary resource for quantum state preparation.

Realistic quantum annealers are open systems, and their interaction with the environment inevitably induces dissipation, which can disrupt adiabatic evolution~\cite{SarandyLidar2005,Amin2008,Dickson2013,Amin2015,Marshall2017,Mishra2018,AlbashLidar2015}.
In this work, we propose an engineered dissipative protocol that helps speed up adiabatic state preparation.
Dissipation alone has been explored to prepare quantum states.
For example, reservoir engineering enables stabilization of target states~\cite{Poyatos1996,Kraus2008,Diehl2008,Verstraete2009,harrington2022engineered}, with applications to correlated many-body systems demonstrated experimentally~\cite{mi2024stable} and investigated through quasiparticle cooling~\cite{Lloyd2025QuasiparticleCooling}.
Algorithmic developments include Lindbladian ground-state preparation~\cite{ding2024single,li2025dissipative,zhan2026rapid} and system-bath protocols for thermal and ground-state preparation with end-to-end efficiency guarantees for specific Hamiltonian classes~\cite{Ding2026EndToEnd}.
Related algorithms address dissipative search~\cite{PhysRevA.111.042430} and local energy minimization under thermal perturbations~\cite{Chen2025}.
Complementary approaches combine cooling with coherent dynamics~\cite{Feng2022CoherentCooling,feng2024escaping} or incorporate dissipation into continuation~\cite{watts2026rapid}, quantum annealing~\cite{Marshall2019PowerPausing,Chen2020WhyPausing,passarelli2020reverse,najafabadi2023improving}, and adiabatic transfer~\cite{menu2022reservoir,King2024AdiabaticTrajectories}.
For dissipation-assisted adiabatic preparation, a central issue is how relaxation modifies nonadiabatic leakage and the minimum-gap dependence of the preparation time.

Our protocol introduces a spectrally filtered dissipative channel in the instantaneous energy eigenbasis. In the ideal downward-only limit, the engineered jump operator induces transitions from higher-energy eigenstates to lower-energy ones while leaving the instantaneous ground state dark. During the annealing evolution, this channel continuously relaxes excitations generated by nonadiabatic driving, helping the system track its instantaneous ground state.

We analyze this mechanism first in an effective two-level avoided-crossing model. Within the effective time-local Lindblad model, when the engineered relaxation rate is large compared with the minimum gap, the dissipative channel suppresses the excited-state population and changes the gap dependence of the required runtime from the closed-system behavior $O(\Delta^{-2})$ to $O(\Delta^{-1})$. For the standard unstructured-search gap scaling, this inverse-gap dependence recovers Grover's square-root runtime scaling within the effective model, at fixed effective relaxation rate and target accuracy. The microscopic conditions under which this effective strong-relaxation regime can be realized are discussed in Sec.~\ref{Physical implementation}. We then include finite-temperature upward transitions satisfying detailed balance and show that they introduce an intrinsic thermal error floor. The inverse-gap estimate remains applicable when the thermal excited-state population at the minimum gap stays below a fixed fraction of the target error tolerance.

The remainder of the paper is outlined as follows. In Sec.~\ref{Model}, we introduce the adiabatic state-preparation model and construct the filtered dissipative channel in the instantaneous energy basis. In Sec.~\ref{Finite-temperature perturbations}, we analyze finite-temperature perturbations and the resulting thermal error floor. In Sec.~\ref{Numerical results}, we present numerical simulations for transverse-field Ising chains, random spin-glass instances, different coupling operators, and finite heating. In Sec.~\ref{Physical implementation}, we discuss a possible superconducting-circuit implementation based on structured bosonic reservoirs and Purcell-filtered microwave environments. We summarize our results in Sec.~\ref{conclusion}.

\section{Model} 
\label{Model}
\subsection{Adiabatic evolution and nonadiabatic leakage}
In standard adiabatic quantum computation~\cite{Farhi2000,Kadowaki1998,AlbashLidar2018}, the system is initialized in the easily prepared ground state of a transverse-field Hamiltonian $H_b$, and the system slowly evolves to a problem Hamiltonian $H_p$ whose ground state encodes the solution to the computational problem. Thus the adiabatic Hamiltonian, 
\begin{equation}
    H(s)=(1-s)H_{b}+sH_{p}, \quad s = \frac{t}{T} \in [0, 1]
\end{equation}
where $T$ is the total annealing time. The instantaneous eigenvalue equation is given by $H(s)|n(s)\rangle = E_n(s)|n(s)\rangle$. 

In a closed system, nonadiabatic transitions are generated by the rotation of the instantaneous eigenbasis. Near a narrow avoided crossing, the dynamics can be approximated by an effective two-level Landau-Zener (LZ) Hamiltonian~\cite{Landau1932,Zener1932}
\begin{equation}
    H_\text{LZ}(t) = \frac{1}{2}[\epsilon(t) \sigma_z + \Delta \sigma_x]
\end{equation}
where $\Delta$ is the minimum gap and $\epsilon(t) = v(\frac{t}{T} - s_*)$ goes to zero at the avoided crossing($t = t_*$, with $s_* = t_*/T$). In the instantaneous eigenbasis, this Hamiltonian becomes
\begin{equation}
    H_\text{ad}(t) = \frac{\Omega(t)}{2} \sigma_z - \frac{\dot{\theta}(t)}{2} \sigma_y 
\end{equation}
where $\Omega(t) = \sqrt{\epsilon^2(t) + \Delta^2}$ and $\tan \theta(t) = \Delta / \epsilon(t)$.  The second term in $H_\text{ad}$ represents the nonadiabatic coupling, which coherently transfers population from the instantaneous ground state to the excited state, thereby limiting the speed of adiabatic state preparation. Suppressing this maximum excitation probability to within a small error tolerance $\varepsilon$ requires a time complexity of $\mathcal{O}(\Delta^{-2})$~\cite{Kato1950,AlbashLidar2018}.

\subsection{Engineered dissipative channel and filtered jump operator}
To suppress nonadiabatic leakage during the adiabatic evolution, we introduce an engineered dissipative channel that relaxes excitations in the instantaneous energy eigenbasis. The system dynamics is described by the Lindblad master equation
\begin{equation}
    \dot{\rho}(s)=-i[H(s),\rho]+\Gamma\mathcal{D}[K(s)]\rho(s)
    \label{eq4}
\end{equation}
where $\mathcal{D}[K]\rho = K\rho K^\dagger - \frac{1}{2}\{K^\dagger K, \rho\}$ is the standard Lindblad dissipator, $\Gamma$ denotes the overall strength of the engineered reservoir and $K(s)$ is the engineered jump operator. The central requirement is that $K(s)$ should induce transitions only from higher-energy instantaneous eigenstates to lower-energy instantaneous eigenstates. Therefore, in the instantaneous eigenbasis of $H(s)$, we choose~\cite{ding2024single, zhan2026rapid}
\begin{equation}
    K(s) = \sum_{m<n} \kappa_{mn}(s)\ket{m(s)}\bra{n(s)}
    \label{eq5}
\end{equation}
where $\kappa_{mn}(s) = \hat{f}(E_m(s) - E_n(s)) \bra{m(s)} A \ket{n(s)}$ with $E_m(s) < E_n(s)$ and $\hat{f}(\omega)$ is a filter function in the frequency domain. This form describes a directional cooling process in the instantaneous eigenbasis. Population that is transferred to excited states by nonadiabatic driving is continuously relaxed back toward lower-energy states. We emphasize that Eq.~\eqref{eq4} is an effective time-local master equation whose dissipator is constructed from the instantaneous eigenstates of the annealing Hamiltonian. Although the laboratory-frame coupling operator $A$ and the filter spectrum $\hat f(\omega)$ are fixed, the jump operator $K(s)$, and hence the effective dissipator, depends parametrically on the annealing control through the instantaneous eigenstates and transition frequencies of $H(s)$. This differs from the fixed-dissipator approximation, in which the dissipator derived for the undriven bare Hamiltonian is kept unchanged after the control Hamiltonian is applied~\cite{Giorgi2020Driven}.

A key property of this construction is that the instantaneous ground state is a dark state of the dissipator. Since there is no state below $\ket{0(s)}$, one has
\begin{equation}
    K(s) \ket{0(s)} =0.
\end{equation}
Consequently,
\begin{equation}
    \mathcal{D}[K(s)](\ket{0(s)} \bra{0(s)}) = 0. 
\end{equation}
Thus, the engineered reservoir does not directly disturb the desired state. Its role is not to dephase the instantaneous ground state, but to remove population that has leaked into excited states. In the absence of nonadiabatic leakage, the ideal engineered channel has no population dynamics to correct. It becomes operative only after finite-rate driving transfers population out of the instantaneous ground state.

It is worth noting that while the filtered jump operator $K(s)$ in Eq.~\eqref{eq5} is formulated explicitly using the instantaneous energy eigenstates $|n(s)\rangle$, this spectral-domain representation does not imply that an experimental system must trace or diagonalize the many-body Hamiltonian in real-time. Instead, this idealized $K(s)$ acts as a mathematically elegant abstraction of the effective secular dynamics of a physical system coupled to a structured environment in the time domain. As will be derived microscopically in Sec.~\ref{Physical implementation}, when a system interacts with a tailored reservoir via a time-independent physical operator, the time-domain bath correlation automatically performs a natural Fourier filtering. Under the Born-Markov and secular approximations, this interaction selectively retains downward transitions while suppressing upward ones, providing a microscopic motivation for the basis-projected operator $K(s)$ used in our numerical analysis under the stated approximations.

\subsection{Effective two-level dynamics near the minimum gap}
\label{Model3}
In the two-level subspace near the minimum gap, the dissipative dynamics reduces to
\begin{equation}
    \dot{\rho}(s)=-i[H_\text{ad}(s),\rho]+\Gamma\mathcal{D}[\sigma^-]\rho(s)
\end{equation}
where $\sigma^- = \ket{0(s)}\bra{1(s)}$ and $\Gamma$ is the engineered relaxation rate from the first excited state to the instantaneous ground state. The populations and coherences then obey
\begin{equation}
    \begin{aligned}
        \dot{p} &= -\Gamma p + \frac{\dot{\theta}}{2} (c + c^*) \\ 
        \dot{c} &= -\left( i\Omega + \frac{\Gamma}{2} \right) c + \frac{\dot{\theta}}{2}(1 - 2p)
    \end{aligned}
\end{equation}
where $p = \rho_{11}$ is the excited-state population and $c = \rho_{01}$ is the coherence. For slow variation of the instantaneous basis, the coherence approximately follows the population dynamics adiabatically. In the regime $p \ll 1$, 
\begin{equation}
    c \approx \frac{\dot{\theta}}{2(i\Omega + \Gamma / 2)}
\end{equation}
The resulting quasi-steady excited-state population is
\begin{equation}
    p(t) \approx \frac{\dot{\theta}^2}{4\Omega^2(t) + \Gamma^2}.
\end{equation}
At the minimum gap $\Omega = \Delta$, this becomes
\begin{equation}
    p_\text{max} = \frac{\dot{\theta}^2}{4\Delta^2 + \Gamma^2}.
\end{equation}
By bounding the maximum excitation probability below a small error tolerance $\varepsilon$, we obtain a lower bound on the total annealing time:
\begin{equation}
    T \geq \frac{v}{\sqrt{\varepsilon} \Delta \sqrt{4\Delta^2 + \Gamma^2}}
    \label{eq13}
\end{equation}
Therefore, engineered dissipation suppresses the conversion of nonadiabatic coherence into excited-state population. For $\Gamma=0$, Eq.~\eqref{eq13} recovers the closed-system scaling $T=O(\Delta^{-2})$. For any $\Gamma>0$, the factor $\sqrt{4\Delta^2+\Gamma^2}$ in the denominator increases, thereby reducing the bound on the required annealing time. In particular, in the strong-dissipation regime $\Gamma\gg\Delta$, Eq.~\eqref{eq13} gives $T=O\!\left((\Gamma\Delta)^{-1}\right)$. For fixed $\Gamma$, this corresponds to the improved small-gap dependence $T=O(\Delta^{-1})$, instead of the closed-system scaling $T=O(\Delta^{-2})$.

We emphasize that this inverse-gap scaling is distinct from the $T\sim \Delta^{-1}$ scaling obtained in adiabatic quantum search~\cite{Roland2002QuantumSearch}. In adiabatic search, the improvement over the naive closed-system $T\sim \Delta^{-2}$ estimate arises from a coherent local adiabatic schedule that uses the known gap profile of the special two-dimensional search Hamiltonian, slowing the evolution only near the minimum gap. By contrast, the present mechanism does not rely on a problem-specific optimized schedule or on the coherent amplification structure of Grover search. Instead, the engineered reservoir changes the local nonadiabatic response itself: nonadiabatic coherence is continuously converted into downward relaxation while the instantaneous ground state remains dark. Thus the same inverse-gap scaling reflects an open-system cooling mechanism rather than a closed-system local-schedule optimization.

While the two-level model provides a clear picture of the dissipative suppression of nonadiabatic leakage, its extension to a multilevel system requires an additional connectivity condition. Specifically, every excited state appreciably populated by the nonadiabatic driving must possess a downward relaxation path to the instantaneous ground-state sector, with an effective rate that does not vanish as the minimum gap decreases. In Appendix~\ref{appendixA}, we derive the corresponding population dynamics and show that the $T=O(\Delta_{n0}^{-1})$ gap dependence extends to the multilevel setting under this condition. Dark or parametrically weakly connected excited states fall outside this scaling regime.

\section{Finite-temperature perturbations}
\label{Finite-temperature perturbations}
In ideal closed-system adiabatic quantum computation, the main obstruction to ground-state preparation is the nonadiabatic transition induced by the finite minimum gap. In realistic quantum annealers, however, the evolution is neither perfectly isolated nor at zero temperature. The operating temperature can be comparable to, or even much larger than, the minimum gap of the problem Hamiltonian. In this regime, the success probability is not determined solely by the minimum gap, but also by the number of thermally accessible excited states near the critical region. This effect can lead to behavior that is qualitatively different from the closed-system prediction, including non-monotonic success probabilities and thermally assisted relaxation processes~\cite{Amin2008,Dickson2013,Mishra2018}.

Another important finite-temperature effect is freeze-out. For sufficiently long annealing times, an open quantum annealer may enter a quasi-static regime in which the instantaneous populations approximately follow a Boltzmann distribution. The final output distribution is then often close to the equilibrium distribution at a freeze-out point, rather than the ground state of the final Hamiltonian. Therefore, making the annealing time arbitrarily long does not necessarily improve the ground-state probability in an open system~\cite{Amin2015,Marshall2017}. This is in sharp contrast to the closed-system adiabatic picture, where slower evolution is always beneficial in principle.

The role of decoherence also depends on the system-bath coupling regime. In the weak-coupling limit, decoherence acts primarily in the instantaneous energy eigenbasis, and a short single-qubit dephasing time does not by itself imply the failure of adiabatic quantum computation. The dominant harmful process is instead thermal excitation out of the instantaneous ground state. In contrast, if decoherence acts in the computational basis, it can destroy the coherence of the instantaneous eigenstates and drive the system toward a mixed state~\cite{AlbashLidar2015}. These observations motivate treating finite temperature not as a small technical correction, but as a fundamental perturbation that competes with adiabatic ground-state preparation.

Motivated by these considerations, we now analyze how finite-temperature perturbations modify the engineered dissipative protocol. In realistic experimental scenarios, the engineered bath is maintained at a finite temperature $T_B>0$, with inverse temperature $\beta=1/k_B T_B$. Consequently, the ideal one-sided filter condition $\hat f(\omega\ge 0)=0$ is relaxed, and thermal fluctuations inevitably induce undesired upward transitions from the instantaneous ground state to excited states.

To accurately capture these non-equilibrium dynamics within the two-level subspace near the minimum gap, the master equation must be generalized to include the thermal excitation channel:
\begin{equation}
    \dot{\rho}=-i[H_{ad}(t),\rho]+\Gamma\mathcal{D}[\sigma^{-}]\rho+\gamma\mathcal{D}[\sigma^{+}]\rho
\end{equation}
where $\gamma$ is the thermal excitation rate driven by the jump operator $\sigma^+ = |1(s)\rangle\langle 0(s)|$. According to the fluctuation-dissipation theorem and the detailed balance condition, the ratio of the upward to downward transition rates is strictly governed by the thermal Boltzmann factor:
\begin{equation}
    \frac{\gamma}{\Gamma}=e^{-\beta\Omega(t)}
\end{equation}
Here $\Gamma(t)$ and $\gamma(t)$ are instantaneous rates, and their time dependence is suppressed below for compactness. For a bath at fixed inverse temperature $\beta$, their ratio therefore varies with the instantaneous gap. A thermal many-body extension requires frequency-resolved jump channels satisfying $W_{n\leftarrow m}(s)/W_{m\leftarrow n}(s) =e^{-\beta[E_n(s)-E_m(s)]}$ for each allowed transition with $E_n(s)>E_m(s)$, where $W_{n\leftarrow m}(s)$ denotes the transition rate from $m$ to $n$. $\Omega(t) = \sqrt{\epsilon(t)^2 + \Delta^2}$ is the instantaneous energy gap. In the quasi-steady-state regime ($\dot{c} \approx 0$) and assuming slow evolution, the adiabatic elimination of the coherence yields the modified population dynamics:
\begin{equation}
    \dot{p}=-(\Gamma+\gamma)p+\gamma+(\Gamma+\gamma)A(t)(1-2p)
\end{equation}
where $A(t)=\frac{\dot{\theta}^2}{4\Omega^2+(\Gamma+\gamma)^2}$ represents the effective non-adiabatic driving. By setting $\dot{p} \approx 0$ and expanding in the weak-driving limit ($A \ll 1$), the quasi-steady population near the minimum gap evaluates to:
\begin{equation}
    p\approx\frac{\gamma}{\Gamma+\gamma}+\left(1-2\frac{\gamma}{\Gamma+\gamma}\right)\frac{v^2}{(4\Delta^2+(\Gamma+\gamma)^2)T^2\Delta^2}
    \label{eq17}
\end{equation}
The first term is the thermal excited-state population at the minimum-gap point,
\begin{equation}
    p_{\mathrm{th}}(\Delta)=\frac{\gamma}{\Gamma+\gamma} =\frac{1}{e^{\beta\Delta}+1},
\end{equation}
where the transition rates are evaluated at $\Omega=\Delta$. The second term describes the nonadiabatic correction within the quasi-steady-state and weak-driving approximations. For a target excitation probability $0<\varepsilon<1/2$, the present tracking criterion requires $p_{\mathrm{th}}<\varepsilon$, or equivalently,
\begin{equation}
    \beta\Delta>\ln\frac{1-\varepsilon}{\varepsilon}.
\end{equation}Since \(0\leq1-2p_{\mathrm{th}}\leq1\), replacing the prefactor of the nonadiabatic term in Eq.~\eqref{eq17} by unity gives a conservative sufficient runtime estimate within these approximations:
\begin{equation}
    T\ge\frac{v}{\Delta\sqrt{4\Delta^2+\Gamma^2(1+e^{-\beta\Delta})^2}\sqrt{\varepsilon-\frac{1}{e^{\beta\Delta}+1}}}
    \label{eq18}
\end{equation}
To make the gap dependence explicit, we allocate a fixed fraction of the error tolerance to thermal excitation by requiring
\begin{equation}
    p_{\mathrm{th}}(\Delta)\leq\alpha\varepsilon,
\qquad 0<\alpha<1,
\end{equation}
where $\alpha$ is independent of $\Delta$. This condition is equivalent to
\begin{equation}
    \beta\Delta\geq \ln\frac{1-\alpha\varepsilon}{\alpha\varepsilon},
\end{equation}
and ensures that $\varepsilon-p_{\mathrm{th}}\geq(1-\alpha)\varepsilon$. Using this inequality and $1+e^{-\beta\Delta}\geq1$ in Eq.~\eqref{eq18}, a conservative sufficient choice of runtime is
\begin{equation}
    T\geq\frac{v}{\Delta\sqrt{4\Delta^2+\Gamma^2}\sqrt{(1-\alpha)\varepsilon}}.
\end{equation}
In the strong-relaxation regime $\Gamma\gg\Delta$, this estimate has the dependence $T=O((\Gamma\Delta)^{-1})$. For fixed $\Gamma$, $v$, $\varepsilon$, and $\alpha$, it therefore retains the inverse-gap dependence $T=O(\Delta^{-1})$, provided that the temperature condition above is maintained and the effective-model approximations remain valid.
At fixed nonzero bath temperature, however, $p_{\mathrm{th}}\rightarrow1/2$ as $\Delta\rightarrow0$, and this condition eventually fails. Thus, at fixed temperature, the result applies only over the gap range satisfying the stated error-budget condition. An asymptotic inverse-gap scaling at fixed target accuracy requires the bath temperature to decrease with the minimum gap; it is not established here in the limit $\Delta\rightarrow0$ at fixed nonzero temperature.

\section{Numerical results}
\label{Numerical results}
We numerically test the engineered dissipative protocol in finite-size many-body systems. 
The main quantity used to characterize the performance is the final ground-state probability
\begin{equation}
P_{\rm GS}(T)= {\rm Tr}\left[\Pi_0\,\rho(T)\right],
\end{equation}
where $\Pi_0$ is the projector onto the ground-state manifold of the final problem Hamiltonian.  
For comparison, we simulate both the closed-system and the dissipative annealing dynamics. The closed-system evolution is governed by the Schr\"odinger equation under the time-dependent Hamiltonian $H(s)$. For the dissipative protocol, the nonadiabatic leakage primarily occurs between the lowest-lying states near the minimum gap, and the engineered downward relaxation actively prevents population from climbing into highly excited states. Therefore, it is physically justified and computationally highly efficient to project the Lindblad dynamics onto a truncated subspace spanned by the lowest instantaneous eigenstates. This approximation accurately captures the dominant nonadiabatic and relaxation processes at the low-energy bottleneck, effectively overcoming the exponential scaling of the full $2^N \times 2^N$ density matrix.
\subsection{Transverse-field Ising model (TFIM)} 
\begin{figure*}[t]
\centerline{
\includegraphics[width=1\textwidth]{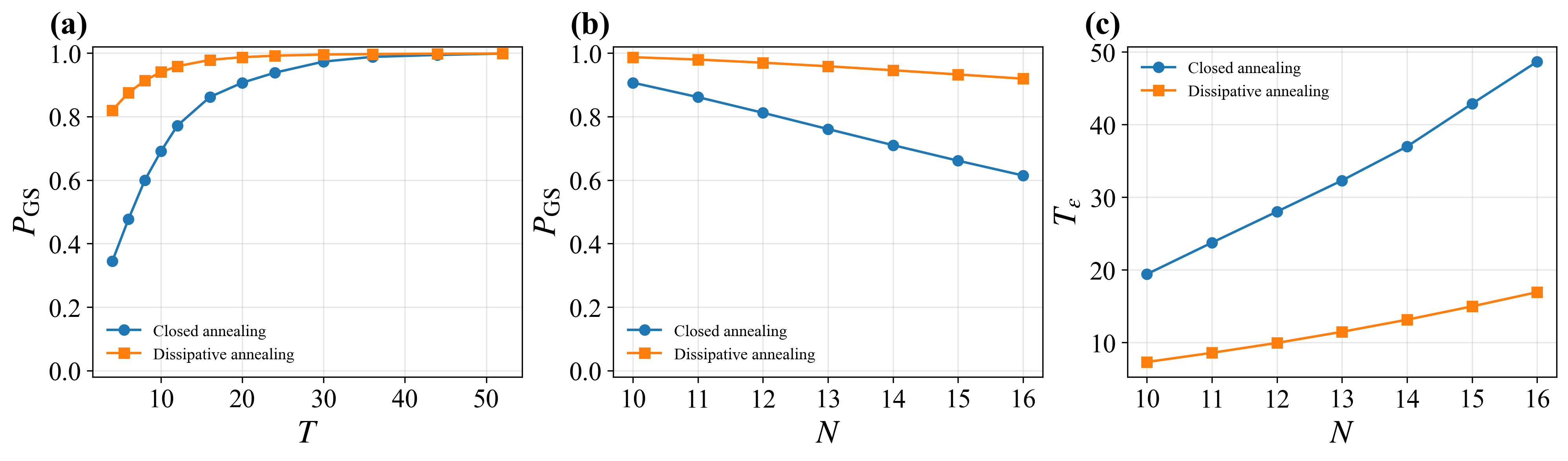}}
\caption{
Numerical results for TFIM. 
(a) Final ground-state probability $P_{\rm GS}$ as a function of the total annealing time $T$ for $N=10$. 
The dissipative protocol reaches high success probability at shorter annealing times than closed-system annealing. 
(b) $P_{\rm GS}$ as a function of system size $N$ at fixed $T=20$. 
The closed-system success probability decreases with increasing $N$, while the dissipative protocol remains more robust. 
(c) Threshold annealing time $T_\varepsilon$ required to reach $P_{\rm GS}\geq0.90$. 
The engineered dissipative protocol substantially reduces the required runtime over the simulated system sizes. 
We set $J=1$, $h=0.3$, and $\Gamma=3$.
}
\label{fig1}
\end{figure*}
The TFIM Hamiltonian with $N$ sites reads:
\begin{equation}
H_p=-J\sum_{i=1}^{N-1}Z_iZ_{i+1}-h\sum_{i=1}^{N}X_i .
\end{equation}
where $J$ is the interaction coupling, $h$ is the transverse field strength, and $Z_{i}$, $X_{i}$ are Pauli operators for the $i$-th site. The dimension of the Hilbert space is $2^{N}$. In our simulations, we set $J=1$ and $h=0.3$.

The system is initialized in the ground state of the transverse-field Hamiltonian $H_{b} = -\sum_{i=1}^{N} X_{i}$, namely the product state $|+\rangle^{\otimes N}$. In the dissipative simulation, the engineered jump operator is constructed in the instantaneous eigenbasis of $H(s)$ by keeping only the downward matrix elements of the operator $A=\partial_s H=H_p-H_b$.

Fig.~\ref{fig1}(a) shows the final ground-state probability $P_{GS}$ as a function of the total annealing time $T$ for a representative chain size $N=10$, where the engineered relaxation rate is set to $\Gamma=3$.
For short and intermediate annealing times, the dissipative protocol gives a substantially larger success probability than closed-system annealing. 
This behavior is consistent with the physical picture developed above: nonadiabatic driving transfers population out of the instantaneous ground state, while the engineered reservoir continuously relaxes this leaked population back toward lower-energy instantaneous eigenstates. 
At sufficiently long annealing times, both protocols approach high ground-state probability, but the dissipative protocol reaches this regime at a much shorter runtime.

Fig.~\ref{fig1}(b) compares the system-size dependence at fixed annealing time $T = 20$. 
As $N$ increases, the closed-system success probability decreases significantly, reflecting the increasing difficulty of maintaining adiabaticity in larger systems. 
By contrast, the dissipative protocol remains much more robust over the same range of system sizes. 
This indicates that the engineered relaxation channel partially compensates for the enhanced nonadiabatic leakage caused by the shrinking low-energy gaps.

To quantify the runtime advantage more directly, Fig.~\ref{fig1}(c) plots the threshold annealing time $T_\varepsilon$, defined as the smallest annealing time for which
\begin{equation}
P_{\rm GS}(T_\epsilon)\geq 1-\varepsilon .
\end{equation}
In our simulations we take $1-\varepsilon=0.90$. 
The required $T_\varepsilon$ grows with system size for both protocols, but the dissipative protocol consistently requires a much shorter annealing time than the closed-system protocol. 
This finite-size result supports the analytic prediction that engineered relaxation suppresses nonadiabatic excitation and reduces the runtime needed to achieve a fixed target success probability.

Overall, the TFIM simulations demonstrate that the directional dissipative channel improves ground-state preparation in three complementary senses: it increases $P_{\rm GS}$ at fixed $T$, improves robustness against increasing system size, and lowers the runtime threshold $T_\varepsilon$ required to reach a prescribed success probability.

\subsection{Spin glass model} 
\begin{figure}[htbp]
\centerline{
\includegraphics[width=0.44\textwidth]{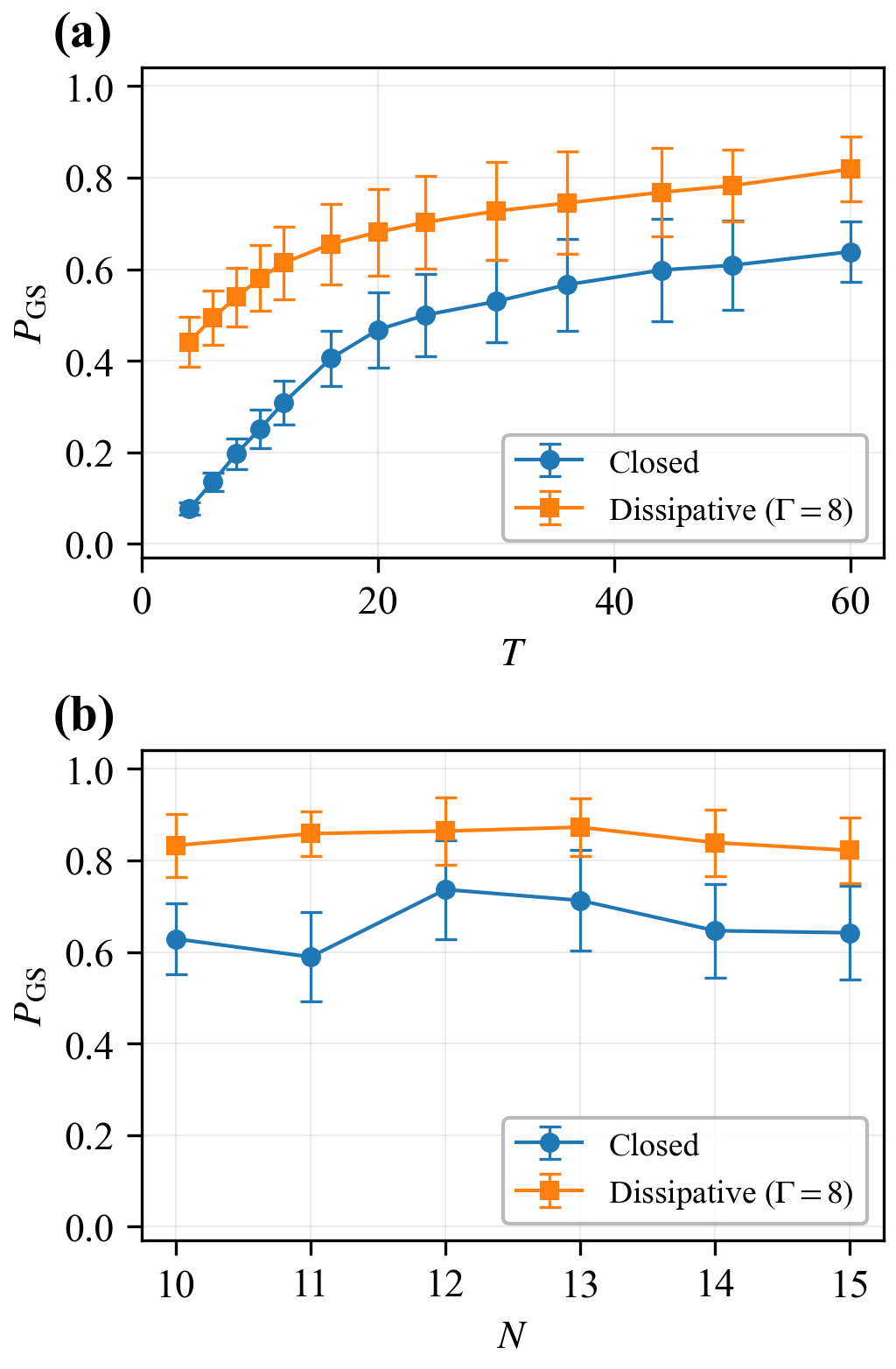}}
\caption{
Numerical results for random spin-glass instances. (a) Disorder-averaged final ground-state probability $P_{GS}$ as a function of the total annealing time $T$ for system size $N= 10$. The dissipative protocol reaches a higher success probability than closed-system annealing over the simulated time range. (b) Disorder-averaged $P_{GS}$ as a function of system size $N$ at a fixed annealing time of $T=60$. In both panels, each data point represents an average over 50 independent random disorder instances, and the error bars indicate the standard deviation. The engineered downward relaxation rate is set to $\Gamma=8$.}
\label{fig2}
\end{figure}
We next test the engineered dissipative protocol on random spin-glass instances. Compared with the uniform TFIM, the spin-glass model has a more irregular low-energy spectrum and stronger sample-to-sample fluctuations, and therefore provides a more stringent test of whether the dissipative mechanism remains useful beyond translationally invariant systems. We consider a spin-glass problem Hamiltonian described by
\begin{equation}
    H_p=-\sum_{i<j}J_{ij} Z_i Z_{j}-\sum_{i=1}^{N}h_i Z_i ,
\end{equation}
where the couplings $J_{ij}$ and local longitudinal fields $h_{i}$ are independently sampled from normal distributions. For each system size, the data are averaged over 50 independent random disorder instances. The dissipative jump operator is constructed in the same way as in the TFIM simulations.

Fig.~\ref{fig2}(a) shows the disorder-averaged final ground-state probability $P_{GS}$ as a function of the total annealing time $T$, where the downward relaxation rate is set to $\Gamma=8$. The dissipative protocol gives a clearly larger success probability than closed-system annealing over the simulated time range. This indicates that the engineered relaxation channel can efficiently remove population that is nonadiabatically transferred into excited instantaneous eigenstates, even when the low-energy spectrum is disordered and contains irregular avoided crossings.

Fig.~\ref{fig2}(b) shows the disorder-averaged system-size dependence of $P_{GS}$ at a fixed annealing time. Despite the pronounced sample-to-sample fluctuations inherent in individual instances, the ensemble average over  disorder realizations confirms that the dissipative protocol consistently outperforms the closed-system baseline. This behavior suggests that the improvement observed in the TFIM is not a consequence of translation invariance or integrability. Instead, the same physical mechanism applies to disordered systems: nonadiabatic leakage is continuously redirected toward lower-energy instantaneous eigenstates by the engineered reservoir.

Overall, the spin-glass simulations support the generality of the dissipative acceleration mechanism. Even in the presence of disorder and sample-dependent spectral bottlenecks, the engineered downward relaxation channel increases the final ground-state probability and improves the robustness of adiabatic state preparation.

\subsection{Influence of different coupling operators} 
\begin{figure}[htbp]
\centerline{
\includegraphics[width=0.44\textwidth]{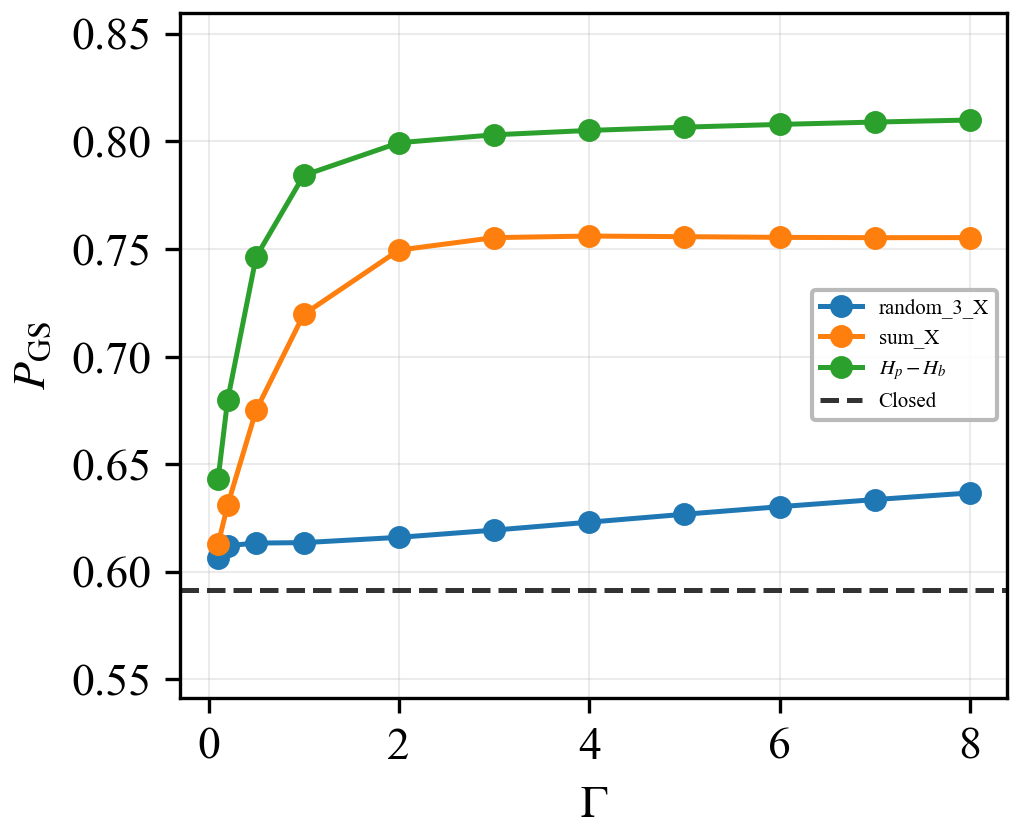}}
\caption{
Disorder-averaged final ground-state probability $P_{GS}$ as a function of the engineered downward relaxation strength $\Gamma$ for different system-bath coupling operators $A$. The performance is evaluated on random spin-glass instances with system size $N=10$ at a fixed annealing time $T=50$. Three coupling configurations are compared: the full adiabatic deformation $A = H_p - H_b$ (green dots), a globally weighted sum of local transverse fields $A = \sum_i w_i X_i$ (orange dots), and a sparse local operator $A = \sum_{i \in \mathcal{S}} X_i$ acting on three randomly selected sites (blue dots). The black dashed line denotes the closed-system baseline. Each data point represents an average over 10 independent disorder realizations.
}
\label{fig3}
\end{figure}
In the preceding simulations, the engineered jump operator was constructed using the ideal adiabatic deformation operator, $A = H_p - H_b$. The choice \(A=\partial_sH\) should therefore be understood as an ideal benchmark that maximizes the overlap between the dissipative transitions and the leakage channels generated by the adiabatic driving. It is not intended to imply that this coupling is generically easier to implement than a dissipator designed for stand-alone ground-state preparation. In practical physical implementations, however, the available system-bath coupling operators are often constrained by the specific hardware architecture, typically limiting the control to local or restricted interactions. To investigate how the choice of the coupling operator $A$ affects the efficiency of the dissipative speedup, we compare the performance of three distinct forms of $A$ as a function of the downward relaxation strength $\Gamma$.

Fig.~\ref{fig3} illustrates the disorder-averaged final ground-state probability $P_{GS}$ for the random spin-glass model under the following coupling configurations: (i) the full adiabatic deformation $A = H_p - H_b$; (ii) a globally weighted sum of local transverse fields, $A = \sum_{i} w_i X_i$, where the weights $w_i$ are normalized and chosen to be proportional to the local longitudinal fields $h_i$; and (iii) a sparse local operator comprising transverse fields acting on only three randomly selected sites, $A = \sum_{i \in \mathcal{S}} X_i$.

Two prominent features emerge from these dynamical results. First, for all three choices of $A$, the ground-state probability $P_{GS}$ initially grows monotonically with the engineered dissipation strength $\Gamma$ before eventually reaching a plateau. This saturation can be competingly understood from the downward-transition network generated by $A$. Increasing $\Gamma$ enhances the rates of transitions with nonzero matrix elements, but does not change the connectivity of the network or remove dark states. Once the well-connected leakage channels have been efficiently relaxed, the residual error is dominated by excited states having small or vanishing matrix elements to lower-energy states. Exact dark states produce a finite error floor, whereas weakly connected states lead to diminishing improvement over the finite range of $\Gamma$ considered here.

Second, the degree of protocol enhancement depends heavily on the spatial structure and support of the 
operator $A$. The global operator $A = H_p - H_b$ yields the highest ground-state preparation fidelity. 
Because nonadiabatic transitions are inherently driven by $\partial_s H$, this choice naturally guarantees 
large matrix elements between the instantaneous ground state and the specific excited states into 
which the population leaks, allowing the engineered reservoir to optimally relax the precise excitations 
generated by the driving fields. The weighted local operator $A = \sum_{i} w_i X_i$ offers intermediate performance, 
demonstrating that hardware-restricted local couplings can still provide substantial acceleration if their spatial 
weights are strategically aligned with the problem's energy landscape. Conversely, the random sparse operator provides only a marginal improvement over the closed-system baseline. This indicates that if $A$ lacks sufficient support across the many-body system, certain nonadiabatic excitations may map into the dark subspaces of the dissipator, thereby escaping the cooling mechanism and degrading the final state preparation fidelity.

\begin{figure*}[t]
\centerline{
\includegraphics[width=1\textwidth]{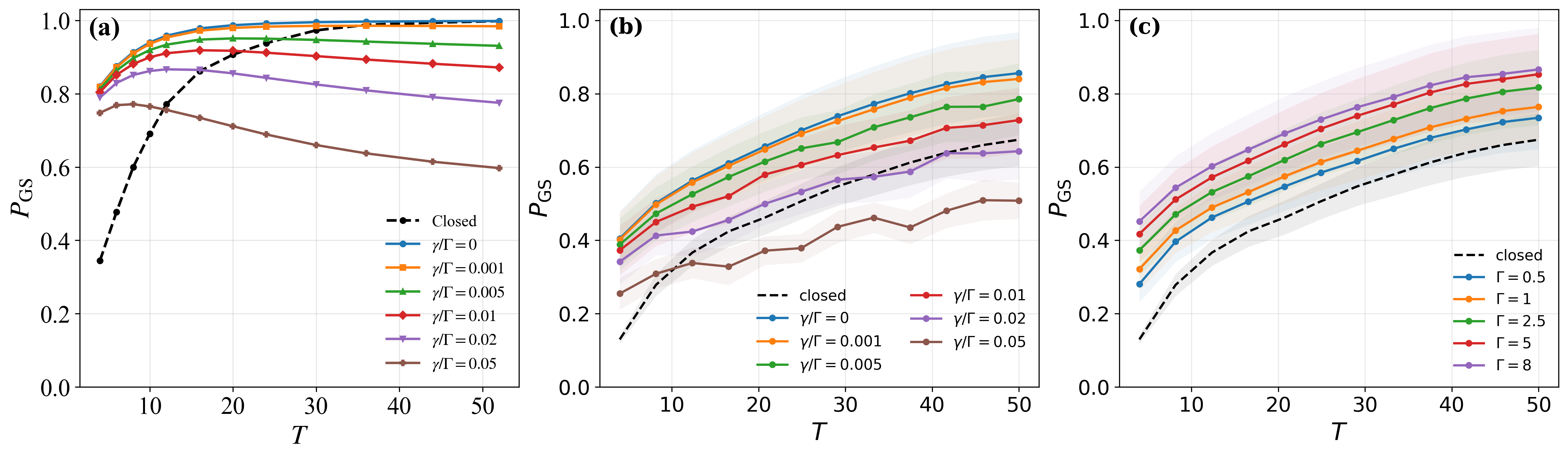}}
\caption{
Robustness of the engineered dissipative protocol against
phenomenological upward-jump errors and the effect of relaxation strength. (a) Final ground-state probability $P_{GS}$ for the TFIM as a function of the total annealing time $T$ for several upward/downward rate ratios $r = \gamma/\Gamma$. The case $r = 0$ corresponds to the ideal zero-temperature filtered reservoir, and the downward relaxation rate is fixed at $\Gamma = 3$. (b) Disorder-averaged $P_{GS}$ for random spin-glass instances as a function of $T$ under the same values of $r$ with $\Gamma = 3$. (c) Disorder-averaged $P_{GS}$ for random spin-glass instances as a function of $T$ for various downward relaxation rates $\Gamma$ (ranging from $0.5$ to $8.0$) under a fixed upward-jump error strength $\gamma = 0.015$. The black dashed line in all panels denotes the closed-system baseline. For both models, the system size is $N = 10$. The spin-glass results in panels (b) and (c) are averaged over 50 independent disorder realizations, and the light shaded regions indicate the standard deviation.
}
\label{fig4}
\end{figure*}
\subsection{Robustness against upward-jump errors}
Finally, motivated by the upward transitions discussed in Sec.~\ref{Finite-temperature perturbations}, we investigate the robustness of the engineered dissipative protocol against imperfect suppression of upward jumps. We introduce a phenomenological upward-jump error channel into the effective Lindblad dynamics:
\begin{equation}
\dot{\rho}
=-i[H(s),\rho] + \Gamma \mathcal{D}[K_{\rm down}(s)]\rho +\gamma\mathcal{D}[K_{\rm up}(s)]\rho ,
\end{equation}
Here $K_{\mathrm{down}}(s)$ is the downward jump operator defined previously, and $K_{\mathrm{up}}(s)=K_{\mathrm{down}}^\dagger(s)$ introduces upward-jump errors. The scalar coefficients $\Gamma$ and $\gamma$, and hence their ratio $r=\gamma/\Gamma$, are held constant during each evolution. Thus, $r$ is a phenomenological error parameter rather than a fixed bath temperature.

The relation to the thermal model of Sec.~\ref{Finite-temperature perturbations} is transparent in the two-level limit. With the Hamiltonian held fixed, the stationary excited-state population is $r/(1+r)$, which equals the thermal population when $r=e^{-\beta\Omega(s)}$. However, a constant $0<r<1$ corresponds to an effective inverse temperature $\beta_{\mathrm{eff}}(s)=-\ln r/\Omega(s)$ that varies along the annealing path. In a many-body system, a single ratio generally cannot satisfy thermal detailed balance at all transition frequencies. Figure~\ref{fig4} therefore tests robustness against upward-jump errors and is not a simulation of a thermal bath at fixed temperature.

Fig.~\ref{fig4}(a) illustrates the final ground-state probability $P_{GS}$ for the TFIM as a function of the annealing time $T$ across several values of $r$. The baseline $r=0$ corresponds to the ideal one-sided filtered reservoir. As expected, increasing $r$ introduces upward-jump errors that partially counteract the engineered relaxation, thereby reducing the final ground-state fidelity. Nevertheless, for sufficiently small $r$, the dissipative protocol continues to achieve a high success probability at short annealing times, significantly outperforming the closed-system evolution. These results demonstrate that the preparation advantage persists under weak upward-jump errors.

Fig.~\ref{fig4}(b) presents analogous results with upward-jump errors for random spin-glass instances. Despite the complex low-energy spectrum and enhanced sample-to-sample fluctuations inherent to the spin-glass model, the qualitative behavior remains consistent: weak upward-jump errors only marginally degrade the dissipative advantage, whereas larger values of $r$ lead to a more pronounced suppression of the preparation fidelity.

To further investigate the competition between engineered relaxation and upward-jump errors, Fig.~\ref{fig4}(c) evaluates the performance of the random spin-glass model for different downward relaxation strengths $\Gamma$, while keeping the upward-jump error strength $\gamma$ fixed. The final ground-state probability increases with $\Gamma$, with diminishing improvement at larger $\Gamma$. Here increasing
$\Gamma$ reduces the relative strength $r=\gamma/\Gamma$ of the independently specified upward-jump error channel. This should be distinguished from increasing the overall coupling to a single thermal bath at fixed temperature, which leaves the detailed-balance ratio, and hence the equilibrium thermal population at a given gap, unchanged.

\section{Physical implementation}
\label{Physical implementation}
We now discuss a possible physical realization of the engineered dissipative channel in a superconducting-circuit platform. The central idea is to couple the annealing system to a structured electromagnetic reservoir whose noise spectrum is strongly asymmetric between upward and downward transitions. Such a construction is closely related to reservoir engineering and to Purcell-filter techniques in circuit quantum electrodynamics (circuit QED)~\cite{Poyatos1996,Verstraete2009,Shankar2013,Blais2004,Blais2021,Reed2010,Sete2015}.

The total Hamiltonian can be written as
\begin{equation}
H_{\rm tot}(t)=H_S(t)+H_B+H_{\rm int},
\end{equation}
where $H_S(t)$ is the annealing Hamiltonian. The engineered bath is modeled as a collection of bosonic microwave modes,
\begin{equation}
H_B=\sum_k \omega_k b_k^\dagger b_k ,
\end{equation}
which may be realized by lossy resonators, transmission lines, or Purcell-filtered microwave environments. The system-bath coupling is taken to be
\begin{equation}
H_{\rm int}=A\otimes \left( \sum_k g_k(b_k+b_k^\dagger)  \right),
\end{equation}
where $A$ is a system operator chosen to have nonzero matrix elements between the relevant instantaneous eigenstates. Therefore, when specifically selecting $A$, it is necessary to ensure that the matrix element $\langle m(s)|A|n(s)\rangle$ has sufficiently large matrix elements near the bottleneck region, while avoiding the introduction of unnecessary dark subspaces.

In the weak-coupling and Markovian regime, the bath enters the system dynamics through its correlation function
\begin{equation}
C(\tau)=\langle B(\tau)B(0)\rangle_B ,
\end{equation}
or equivalently through the noise spectral density
\begin{equation}
S_B(\omega)=\int_{-\infty}^{\infty}d\tau\, e^{-i\omega \tau} C(\tau).
\end{equation}
With the sign convention used here, the transition rate from an instantaneous eigenstate $|n(s)\rangle$ to $|m(s)\rangle$ is proportional to
\begin{equation}
\Gamma_{m\leftarrow n}(s)
=
|\langle m(s)|A|n(s)\rangle|^2
S_B(E_m(s)-E_n(s)).
\end{equation}
Therefore, the mathematical filter function used in the model can be physically interpreted as the bath noise spectrum evaluated at the corresponding transition frequency:
\begin{equation}
\hat f(E_m-E_n)
\ \longleftrightarrow\
S_B(E_m-E_n).
\end{equation}
Although $A$ and $S_B(\omega)$ are fixed in the laboratory frame, the resulting transition rates depend on the annealing control through the instantaneous energy differences $E_m(s)-E_n(s)$ and matrix elements $\langle m(s)|A|n(s)\rangle$. Consequently, the effective dissipator is control dependent. As discussed in Ref.~\cite{Giorgi2020Driven}, a microscopic dissipator for a driven system may reduce to the fixed-dissipator description for an approximately flat environmental spectrum, whereas the two descriptions can differ for a structured reservoir. Our implementation discussion assumes the former instantaneous-eigenbasis construction under the weak-coupling, Markovian, slow-driving, and secular approximations.

For a thermal bosonic reservoir at inverse temperature $\beta=1/k_B T_B$, the spectrum satisfies the detailed-balance relation
\begin{equation}
\frac{S_B(\omega)}{S_B(-\omega)}=e^{-\beta \omega},
\qquad \omega>0 .
\end{equation}
Equivalently, one may write
\begin{equation}
S_B(\omega)=
\begin{cases}
J(|\omega|)\,[1+n_{\rm th}(|\omega|)], & \omega<0,\\[4pt]
J(\omega)\,n_{\rm th}(\omega), & \omega>0,
\end{cases}
\end{equation}
where $J(\omega)$ is the engineered spectral density and the thermal photon occupation $n_{th}(\omega)=(e^{\beta\omega}-1)^{-1}$. In the low-temperature limit $k_{B}T_{B}\ll\Omega$, where $\Omega$ denotes the relevant instantaneous transition energy, upward transitions are exponentially suppressed. The ideal one-sided condition $\hat{f}(\omega\ge0)=0$ is therefore approximately realized.

The remaining task is to shape the magnitude and bandwidth of the downward spectral density $J(\omega)$. In microwave circuits, the relevant spectral density is determined by the admittance or impedance seen by the system operator $A$; schematically, $J(\omega)\propto\omega \text{Re}Y(\omega)$, up to circuit-dependent coupling factors~\cite{Devoret1997,Burkard2004,Clerk2010}. While a standard single-mode lossy resonator or basic Purcell filter provides an experimentally tunable electromagnetic density of states~\cite{Murch2012,Houck2008,Reed2010,Sete2015,Bronn2015}, its fixed, narrow passband presents a challenge for adiabatic annealing. Particularly in complex energy landscapes---such as the random spin-glass instances discussed 
in Sec.~\ref{Numerical results}---the instantaneous transition gaps fluctuate significantly along the annealing trajectory.

To address this and prevent nonadiabatic freeze-out, the system can be coupled to a structured multi-mode environment, 
such as double-mode cavities. Extending the dissipative channel to a multi-mode architecture effectively 
broadens the cooling bandwidth. This configuration acts as a continuous, broadband thermodynamic driver, 
facilitating quantum coherent cooling that allows the system to efficiently mitigate population trapping in 
excited low-energy states across the complex optimization landscape~\cite{feng2024escaping, Bronn2015,Sunada2022,Tan2017}.

This implementation requires several consistency conditions. First, the bath correlation time must be short compared with the annealing timescale, so that the Markov approximation remains valid. Second, the broadened multi-mode filter bandwidth should properly cover the relevant instantaneous transition frequencies during the bottleneck portion of the annealing path. Third, the relaxation rate $\Gamma$ appearing in the effective two-level master equation should be distinguished from the microscopic system-bath coupling strength. Denoting the bath correlation time by $\tau_B$ and the smallest relevant separation between distinct Bohr-frequency sectors by $\delta\omega_{\mathrm{sec}}$, the Born-Markov and secular approximations require~\cite{albash2012quantum}
\begin{equation}
    \Gamma\tau_B\ll1,\qquad \Gamma\ll\delta\omega_{\mathrm{sec}},
\end{equation}
whereas the dissipative speedup requires $\Delta_{\min}\ll\Gamma$. A weak-coupling secular realization therefore requires the hierarchy
\begin{equation}
    \Delta_{\min}\ll\Gamma\ll\min\!\left\{\tau_B^{-1},\delta\omega_{\mathrm{sec}}\right\}.
\end{equation}
Such a separation of scales is system dependent and is not guaranteed for a generic many-body spectrum. 
In particular, when $\delta\omega_{\mathrm{sec}}$ becomes comparable to $\Delta_{\min}$, the conventional 
secular derivation cannot justify the strong-relaxation regime. In that case, the $O(\Delta^{-1})$ result should be understood as a prediction of the effective engineered Lindblad model, while a quantitative microscopic implementation would require a nonsecular treatment. 
Finally, possible Lamb shifts and additional dephasing induced by the structured environment should be calibrated or compensated. Under these conditions, a multi-mode Purcell-filtered superconducting circuit provides a feasible route toward realizing the directional cooling channel assumed in the theoretical model.

\section{conclusion}
\label{conclusion}
In this work, we showed how engineered dissipation can act as a dynamical leakage-correction mechanism during finite-time adiabatic evolution.
The central idea is to use dissipation not as an uncontrolled source of decoherence, but as a designed resource that 
removes nonadiabatic excitations during the annealing process. 
By constructing a filtered jump operator in the instantaneous energy eigenbasis, the dissipative channel induces 
transitions only from higher-energy instantaneous eigenstates to lower-energy ones. 
As a result, the instantaneous ground state remains a dark state of the dissipator, while population leaked into 
excited states by nonadiabatic driving is continuously relaxed back toward the low-energy sector.

Using an effective two-level avoided-crossing model, we showed that the engineered relaxation suppresses the 
buildup of nonadiabatic excitation near the minimum gap. 
In the strong-relaxation regime, this changes the required runtime scaling from the conventional 
closed-system behavior $T=\mathcal{O}(\Delta^{-2})$ to the improved scaling $T=\mathcal{O}(\Delta^{-1})$. 
This result provides a simple physical mechanism for the speedup: the annealing process no longer 
needs to completely avoid excitations, because the engineered reservoir actively removes them as they are generated.

We also analyzed finite-temperature upward transitions, which compete with the engineered downward 
relaxation. Within the effective two-level description, the inverse-gap runtime estimate is retained 
when the thermal population at the minimum gap satisfies $p_{\mathrm{th}}(\Delta)\leq\alpha\varepsilon$, 
with a fixed $0<\alpha<1$, subject to the stated strong-relaxation and effective-model assumptions. 
At fixed nonzero bath temperature, this condition eventually fails as $\Delta\to0$. The finite-temperature result therefore applies within the corresponding
low-temperature gap range and does not establish asymptotic
inverse-gap scaling at fixed nonzero temperature.

Numerical simulations support these analytic conclusions. 
For the transverse-field Ising model, the dissipative protocol achieves larger final ground-state probabilities than closed-system annealing at fixed runtime and requires shorter annealing times to reach a given target success probability. 
For random spin-glass instances, the same qualitative improvement persists after disorder averaging, indicating that the mechanism is not limited to uniform integrable models. 
Separately, simulations with a fixed upward/downward rate ratio demonstrate robustness against weak upward-jump errors.

Finally, we discussed a possible superconducting-circuit implementation based on a structured multi-mode bosonic reservoir and Purcell filtering. In this setting, the filtered jump operator arises from an engineered bath spectral density that is strongly asymmetric between downward and upward transition frequencies. Crucially, to accommodate the dynamically fluctuating energy gaps inherent in complex optimization problems like spin glasses, extending this architecture to multi-mode cavities provides a necessary broadband quantum coherent cooling channel. This active thermodynamic driver prevents nonadiabatic freeze-out and explicitly assists the system in escaping local minima. Although practical implementations must account for residual heating, Lamb shifts, and the validity of weak-coupling and Markovian approximations, the present results suggest that broadband engineered dissipation can serve as a highly controllable resource for faster and more robust adiabatic quantum state preparation.

Our work is complementary to recent dissipative-continuation algorithms for chemical reaction paths~\cite{watts2026rapid}. While those approaches use engineered cooling to propagate warm starts between discrete molecular geometries, the present work uses a filtered reservoir to suppress nonadiabatic leakage during continuous adiabatic evolution. This distinction highlights engineered dissipation as both a continuation stabilizer and a runtime-acceleration mechanism for adiabatic state preparation.

\section{Acknowledgements}
\label{Acknowledgements}
This work is supported by the National Natural Science Foundation of China (Grants No. 92365202, No. 12475011, and No. 11921005), the National Key R\&D Program of China (2024YFA1409002), Shanghai Municipal Science and Technology Major Project (Grant No.2019SHZDZX01).  

\appendix
\section{Derivation of the Dissipative Adiabatic Master Equation}
\label{appendixA}
To generalize the effective two-level dynamics discussed in Sec.~\ref{Model3} and obtain a rigorous bound on the final excitation probability for the full many-body system, we transform the Lindblad master equation into the instantaneous eigenbasis of the system. We define the unitary transformation operator $U(s)=\sum_{n}|n(s)\rangle\langle n(0)|$, The density matrix in the adiabatic frame is given by $\tilde{\rho}=U^{\dagger}(s)\rho U(s)$. Applying this transformation, the master equation becomes:
\begin{equation}
    \partial_{t}\tilde{\rho}=-i[\Lambda(s)+\frac{1}{T}G(s),\tilde{\rho}]+\gamma\mathcal{D}[\tilde{K}(s)]\tilde{\rho}
\end{equation}
where $\Lambda(s)=\sum_{n}E_{n}(s)|n(0)\rangle\langle n(0)|$ is the diagonal matrix of instantaneous eigenvalues, $\tilde{K}(s)=\sum_{m<n}\kappa_{mn}(s)|m(0)\rangle\langle n(0)|$ is the transformed jump operator and $G(s)=iU^{\dagger}(s)\partial_{s}U(s)$ represents the non-adiabatic coupling, with its matrix elements evaluating to:
\begin{equation}
    G_{mn}(s)=i\frac{\langle m(s)|\partial_{s}H(s)|n(s)\rangle}{\Delta_{nm}(s)}
\end{equation}
where $\Delta_{nm}(s) = E_n(s) - E_m(s)$ is the instantaneous energy gap. To understand the population dynamics, we project the master equation onto its diagonal elements $p_{n}=\langle n|\tilde{\rho}|n\rangle$. The unitary part, driven by the non-adiabatic term $G(s)/T$, couples the populations to the off-diagonal coherences $\tilde{\rho}_{kn}$:
\begin{equation}
    \langle n|-i[\Lambda+\frac{1}{T}G,\tilde{\rho}]|n\rangle=-\frac{i}{T}\sum_{k\ne n}(G_{nk}\tilde{\rho}_{kn}-\tilde{\rho}_{nk}G_{kn})
\end{equation}
The dissipative part contributes to the population rate equations through injection and decay terms:
\begin{equation}
    \langle n|\gamma\mathcal{D}[\tilde{K}]\tilde{\rho}|n\rangle = \sum_{m>n}\Gamma_{n\leftarrow m}(s)p_{m}-\sum_{m<n}\Gamma_{m\leftarrow n}(s)p_{n}
\end{equation}
where $\Gamma_{m\leftarrow n}=\gamma|\kappa_{nm}(s)|^{2}$. The coherences $\tilde{\rho}_{kn}$ evolve according to:
\begin{equation}
    \partial_{t}\tilde{\rho}_{kn}=(-i\Delta_{kn}-\frac{\gamma_{kn}}{2})\tilde{\rho}_{kn}-\frac{i}{T}\sum_{j}(G_{kj}\tilde{\rho}_{jn}-\tilde{\rho}_{kj}G_{jn})
\end{equation}
where $\gamma_{kn}/2$ is the coherence decay rate induced by the dissipator. Because the dynamical phases oscillate rapidly compared to the annealing schedule, we can invoke the secular approximation. We set $\partial_{t}\tilde{\rho}_{kn} \approx 0$ and retain only the principal diagonal driving terms, yielding:
\begin{equation}
    \tilde{\rho}_{kn}\approx\frac{-iG_{kn}/T}{i\Delta_{kn}+\gamma_{kn}/2}(p_{n}-p_{k})
\end{equation}
Substituting this quasi-steady-state coherence back into the population equation yields a classical rate equation:
\begin{equation}
    \dot{p}_{n}=\sum_{k\ne n}W_{nk}(p_{k}-p_{n})+\sum_{m>n}\Gamma_{n\leftarrow m}(s)p_{m}-\sum_{m<n}\Gamma_{m\leftarrow n}(s)p_{n}
    \label{eqA7}
\end{equation}
where we have defined the effective non-adiabatic transition rate:
\begin{equation}
    W_{nk}=\frac{1}{T^{2}}\cdot\frac{|\langle n|\partial_{s}H|k\rangle|^{2}\gamma_{kn}}{\Delta_{kn}^{2}(\Delta_{kn}^{2}+(\gamma_{kn}/2)^{2})}
    \label{eqA8}
\end{equation}
We are primarily concerned with the total instantaneous excitation probability $q(t)=1-p_{0}(t)=\sum_{n\ge1}p_{n}(t)$. Assuming a low-excitation limit where $p_0 \approx 1$ and $p_n \ll 1$ for $n \ge 1$ , the dominant non-adiabatic injection from the ground state to the $n$-th excited state is $J_{n}^{na}(s)\approx W_{n0}(s)$. The total excitation rate is bounded by:
\begin{equation}
    \dot{q}(t)\le J^{na}(s)-\lambda_{0}(s)q(t)
\end{equation}
where $J_{\mathrm{na}}(s)=\sum_{n\geq1}J_n^{\mathrm{na}}(s)$ is the total nonadiabatic injection rate and $\lambda_0(s)=\min_{n\geq1}\Gamma_{0\leftarrow n}(s)$ is the smallest total downward relaxation rate among the excited states included in the effective dynamics. Here
\begin{equation}
    \Gamma_{m\leftarrow n}(s)=\Gamma\left|\widehat f(E_m-E_n)\right|^2\left|\langle m(s)|A|n(s)\rangle\right|^2.
\end{equation}
In deriving the following estimate, we impose the stronger sufficient condition that every dynamically relevant excited state has a direct relaxation channel to the instantaneous ground-state sector. Specifically, we assume that $\lambda_0(s)\ge\lambda_*>0$ throughout the annealing path. This condition excludes excited dark states as well as relaxation networks in which population can reach the ground state only through multistep cascades.
\begin{equation}
    q(T)\le e^{-\int_{0}^{T}\lambda_{0}(t)dt}q(0)+\int_{0}^{T}e^{-\int_{t}^{T}\lambda_{0}(t')dt'}J^{na}(t)dt
    \label{eqA12}
\end{equation}
For an initial instantaneous ground state, $q(0)=0$. If $\lambda_0(s)\geq\lambda_*$ and $J_{\mathrm{na}}(s)\leq J_*$, Eq.~\eqref{eqA12} gives
\begin{equation}
    q(T) \leq \frac{J_*}{\lambda_*} \left(1-e^{-\lambda_*T}\right) \leq \frac{J_*}{\lambda_*},
\end{equation} 
where $J_*=\max_sJ_{\mathrm{na}}(s)$. Therefore, the sufficient condition $q(T)\leq\varepsilon$ is $J_*\leq\varepsilon\lambda_*$. Using Eq.~\eqref{eqA8}, this gives
\begin{equation}
    T\geq\frac{1}{\sqrt{\varepsilon\lambda_*}}\left[\max_s\sum_{n\geq1}\frac{\left|\langle n(s)|\partial_sH(s)|0(s)\rangle\right|^2
\gamma_{n0}(s)}{\Delta_{n0}^2(s)\left[\Delta_{n0}^2(s)+\left(\gamma_{n0}(s)/2\right)^2\right]}
\right]^{1/2}.
\end{equation}
In the strong-relaxation regime, $\gamma_{n0}\gg\Delta_{n0}$, this bound has the gap dependence $T=O(\Delta_{\min}^{-1})$, provided that $\lambda_*$, the relevant coherence-decay rates, and the sum of nonadiabatic matrix elements do not introduce additional singular dependence on $\Delta_{\min}$. The inverse-gap scaling is therefore conditional on a connected downward-transition network with nonvanishing effective rates. It does not apply when the nonadiabatically populated subspace contains dark or parametrically weakly connected excited states.

%

\end{document}